\RequirePackage{fix-cm}
\documentclass[smallextended]{svjour3}       
\smartqed  
\usepackage{graphicx}
\journalname{myjournal}

\usepackage{latexsym}
\usepackage{amsmath,amssymb}
\usepackage{stmaryrd}
\usepackage{hyperref}
\usepackage{graphicx}

\usepackage[most]{tcolorbox}

\usepackage[american]{babel}
\usepackage{csquotes}

\usepackage[backend = biber, style = apa, hyperref = true, url = false, doi = true]{biblatex}

\begin{document}

\title{A Simplified Variant of Gödel's Ontological Argument}

\author{
    Christoph Benzmüller
}


\institute{C. Benzmüller \at
              Otto-Friedrich-Universität Bamberg,
              AI Systems Engineering, Bamberg, Germany \\
              Freie Universität Berlin,  Dep. of Mathematics and Computer Science, Berlin, Germany\\
              \email{christoph.benzmueller@uni-bamberg.de} 
              }

\date{}

%





\maketitle

\begin{abstract}
A simplified variant of Gödel's ontological argument is presented. The simplified argument is valid already in basic modal logics K or KT, it does not suffer from modal collapse, and it avoids the rather complex predicates of essence (Ess.) and necessary existence (NE) as used by Gödel. The  variant presented has been obtained as a side result of a series of theory simplification experiments conducted in interaction with a modern proof assistant system. The starting point for these experiments was the computer encoding of Gödel's argument, and then automated reasoning techniques were systematically applied to arrive at the simplified variant presented. The presented work thus exemplifies a fruitful human-computer interaction in computational metaphysics. Whether the presented result increases or decreases the attractiveness and persuasiveness of the ontological argument is a question I would like to pass on to philosophy and theology.
\keywords{Ontological argument \and Computational metaphysics \and Modal collapse}
\end{abstract}

\section{Introduction}
Gödel's (\citeyear{GoedelNotes}) ontological argument has attracted significant, albeit controversial, interest among philosophers, logicians and theologians \parencite{sobel2004logic}. 
In this article I present a simplified variant of Gödel's argument that was developed in interaction with the proof assistant system Isabelle/HOL \parencite{Isabelle}, which is based on classical higher-order logic \parencite{J43}. My personal interest in Gödel's argument has been primarily of logical nature. In particular, this interest encompasses the challenge of automating and applying reasoning in quantified modal logics using an universal meta-logical reasoning approach \parencite{J41} in which (quantified) non-classical logics are semantically embedded in classical higher-order logic.
The simplified ontological argument presented below is a side result of this research, which began with a computer encoding of Gödel's argument so that it became amenable to formal analysis and computer-assisted theory simplification experiments; cf.~\textcite{C85} for more technical details on the most recent series of experiments.  The simplified argument selected for presentation in this article has, I believe, the potential to further stimulate the philosophical and theological debate on Gödel's argument, since the simplifications achieved are indeed quite far-reaching:  

\begin{itemize}
\item Only minimal assumptions about the modal logic used are required.  The simplified variant presented is indeed valid in the comparatively weak modal logics K or KT, which only use uncontroversial reasoning principles.\footnote{Some background on modal logic (see also \cite{sep-logic-modal}, and the references therein): The modal operators $\Box$ and $\Diamond$ are employed, in the given context, to capture the alethic modalities ``necessarily holds'' and ``possibly holds'', and often the modal logic S5 is used for this. However, logic S5 comes with some rather strong reasoning principles, that could, and have been, be taken as basis for criticism on Gödel's argument. Base modal logic K is comparably uncontroversial, since it only adds the following principles to classical logic: (i) If $s$ is a theorem of K, then so is $\boldsymbol{\Box}s$, and (ii) the distribution axiom $\boldsymbol{\Box}(s\boldsymbol{\rightarrow}t)\boldsymbol{\rightarrow}(\boldsymbol{\Box}{s}\boldsymbol{\rightarrow} \Box{t})$ (if $s$ implies $t$ holds necessarily, then the necessity of $s$ implies the necessity of $t$). Modal logic KT additionally provides the T axiom: $\boldsymbol{\Box}s\boldsymbol{\rightarrow}s$ (if $s$ holds necessarily, then $s$), respectively its dual $s\boldsymbol{\rightarrow}\boldsymbol{\Diamond}s$ (if $s$, then $s$ is possible).

Model logics can be given a possible world semantics, so that $\boldsymbol{\Box}s$ can be read as: for all possible worlds $v$, which are reachable from a given current world $w$, we have that $s$ holds in $v$. And its dual, $\boldsymbol{\Box}s$, thus means: there exists a possible world $v$, reachable from the current world $w$, so that $s$ holds in $v$.
}
\item Gödel's argument introduces the comparably complex predicates of essence ($\mathsf{Ess.}$)  and necessary existence ($\mathsf{NE}$), where the latter is based on the former. These terms are avoided altogether in the simplified version presented here.

\item Above all, a controversial side effect of Gödel's argument, the so-called modal collapse, is avoided. 
Modal collapse ($\mathsf{MC}$), formally notated as $\boldsymbol{\forall} s\ (s \boldsymbol{\rightarrow} \boldsymbol{\Box} s)$, expresses that ``what holds that holds necessarily'', which can also be interpreted as ``there are no contingent truths'' and that ``everything is determined''.  The observation that Gödel's argument implies modal collapse has already been made by \textcite{Sobel}, and \textcite{Kovacs2012} argues that modal collapse may even have been intended by Gödel. Indeed, the study of modal collapse has been the catalyst for much recent research on the ontological argument. For example, variants of Gödel's argument that avoid modal collapse have been presented by 
\textcite{Anderson90}, \textcite{AndersonGettings},
and \textcite{fitting02:_types_tableaus_god}, among others, cf.~also the formal verification and comparison of these works by \textcite{J52}. In the following, however, it is shown that modal collapse can in fact be avoided by much simpler means.
\end{itemize}

What I thus present in the remainder is a simple, generalised divine theory, derived from Gödel's argument, that does not entail modal collapse.


Since Gödel's (\citeyear{GoedelNotes}) argument was shown to be inconsistent \parencite{C55}, the actual starting point for the exploration of the simplified ontological argument has been Scott's variant (\citeyear{ScottNotes}), which is consistent. The terminology and notation used in what follows therefore also remains close to Scott's. 

Only one single uninterpreted constant symbol $\mathsf{P}$ is used in the argument. This symbol denotes  ``positive properties'', and its meaning is restricted by the postulated axioms, as discussed below. Moreover, the following definitions (or shorthand notations) were introduced by Gödel, respectively Scott:


\begin{itemize}
\item An entity $x$ is God-like if it possesses all positive properties.
$$\mathsf{G}(x)  \quad \equiv  \quad \boldsymbol{\forall} \phi\ (\mathsf{P}(\phi)
\boldsymbol{\rightarrow} \phi(x))$$
\item A property $\phi$ is an essence ($\mathsf{Ess.}$) of an entity $x$ if, and
only if,  (i) $\phi$ holds for $x$ and (ii) $\phi$ necessarily entails every
property $\psi$ of
$x$ (i.e., the property is necessarily
minimal).
$$\phi\ \mathsf{Ess.}\ x  \quad \equiv  \quad \phi(x) \boldsymbol{\wedge} \boldsymbol{\forall} \psi\  
                  (\psi(x) \boldsymbol{\rightarrow} \boldsymbol{\Box} \boldsymbol{\forall} y\ (\phi(y)
                    \boldsymbol{\rightarrow} \psi(y))) $$

Deviating from Gödel, Scott added here the requirement that $\phi$ must hold for $x$. Scott found it natural to add this clause, not knowing that it fixed the inconsistency in Gödel's theory, which was discovered by an automated theorem prover \parencite{C55}. Gödel's (\citeyear{GoedelNotes}) scriptum avoids this conjunct, although it occurred in some of his earlier notes.
                             
 \item A further shorthand notation, $\mathsf{NE}(x)$, termed necessary existence,  was introduced by Gödel.  $\mathsf{NE}(x)$ expresses that $x$ necessarily exists 
 if all of its essential properties are necessarily exemplified.
 $$\mathsf{NE}(x) \quad \equiv  \quad \boldsymbol{\forall} \phi\ 
                  (\phi\ \mathsf{Ess.}\  x \boldsymbol{\rightarrow}
                  \boldsymbol{\Box} \boldsymbol{\exists}x\ \phi(x))$$
\end{itemize}

The axioms of Scott's (\citeyear{ScottNotes}) theory, which constrain the meaning of constant symbol $\mathsf{P}$, and thus also of definition $\mathsf{G}$, are now as follows:

\begin{description}
\item[\textsf{AXIOM 1}]  
Either a property or its negation is positive, but not both.\footnote{${\boldsymbol{\neg}} \phi$ is shorthand for $\lambda x\ \boldsymbol{\neg}\phi(x)$.}
$$\boldsymbol{\forall} \phi\ ( \mathsf{P}
  ({\boldsymbol{\neg}} \phi)
  \boldsymbol{\leftrightarrow} \boldsymbol{\neg}\mathsf{P}(\phi) )$$
\item[\textsf{AXIOM 2}] 
A property is positive if it is necessarily entailed by a positive property.
 $$\boldsymbol{\forall} \phi\  \boldsymbol{\forall}  \psi\ ((\mathsf{P}(\phi) \boldsymbol{\wedge} (\boldsymbol{\Box} \boldsymbol{\forall} x\ (\phi(x)
                    \boldsymbol{\rightarrow} \psi(x)))) \boldsymbol{\rightarrow} \mathsf{P}(\psi))$$
\item[\textsf{AXIOM 3}] Being Godlike is a positive property.\footnote{Alternatively, we may postulate \textsf{A3'}: The conjunction of any collection of positive properties
is positive. Formally, 
 $\boldsymbol{\forall} \mathcal{Z}. (\mathsf{P}os\ \mathcal{Z}
 \boldsymbol{\rightarrow} \boldsymbol{\forall} X\ (X \boldsymbol{\textstyle\bigsqcap} \mathcal{Z}
 \boldsymbol{\rightarrow} \mathsf{P}\ X))$, where 
 $\mathsf{P}os\ \mathcal{Z}$ stands for $\boldsymbol{\forall} X\ 
                             (\mathcal{Z}\ X \boldsymbol{\rightarrow}
                             \mathsf{P}\ X)$ and  $X \boldsymbol{\textstyle\bigsqcap}  \mathcal{Z}$  is shorthand for 
                              $\boldsymbol{\Box}\boldsymbol{\forall}
                              u. (X\ u \boldsymbol{\leftrightarrow}
                              (\boldsymbol{\forall} Y.\ \mathcal{Z}\ Y
                              \boldsymbol{\rightarrow}  Y\ u))$.}
$$\mathsf{P}(\mathsf{G})$$

\item[\textsf{AXIOM 4}] 
Any positive property is necessarily positive (in Scott's words: being a positive property is logical, hence, necessary).
$$\boldsymbol{\forall} \phi\ (\mathsf{P}(\phi)
  \boldsymbol{\rightarrow} \boldsymbol{\Box}\mathsf{P}(\phi))$$ 
\item[\textsf{AXIOM 5}] 
Necessary existence ($\mathsf{NE}$) is a
positive property.
$$\mathsf{P}(\mathsf{NE})$$ 
\end{description}

From this theory the following theorems and corollaries follow; cf. \textcite{ScottNotes} and \textcite{C40,C55} for further details. 
Note that the proofs are valid already in (extensional) modal logic KB, which extends base modal logic K
with \textsf{AXIOM B}: $\boldsymbol{\forall} s\ 
                  (s \boldsymbol{\rightarrow}
                  \boldsymbol{\Box} \boldsymbol{\Diamond} s)$, or in words, if $s$ then $s$ is necessarily possible.
\begin{description}
\item[\textsf{THEOREM 1}]  Positive properties are possibly exemplified. $$\boldsymbol{\forall} \phi\ 
                  (\mathsf{P}(\phi) \boldsymbol{\rightarrow}
                  \boldsymbol{\Diamond} \boldsymbol{\exists}x\ \phi(x))$$
                  Follows from \textsf{AXIOM 1} and \textsf{AXIOM 2}.
\item[\textsf{CORO}]  Possibly there exists a God-like being. $$\boldsymbol{\Diamond} \boldsymbol{\exists}x\ \mathsf{G}(x)$$
                  Follows from \textsf{THEOREM 1} and \textsf{AXIOM 3}.
\item[\textsf{THEOREM 2}]  Being God-like is an essence of any God-like being.
   $$ \boldsymbol{\forall}x\ \mathsf{G}(x) \boldsymbol{\rightarrow} \mathsf{G}\ \mathsf{Ess.}\ x$$
   Follows from \textsf{AXIOM 1} and \textsf{AXIOM 4} using the definitions of $\mathsf{Ess.}$ and $\mathsf{G}$.
\item[\textsf{THEOREM 3}]  Necessarily, there exists a God-like being. 
   $$\boldsymbol{\Box} \boldsymbol{\exists}x\ \mathsf{G}(x)$$
   Follows from \textsf{AXIOM 5}, \textsf{CORO}, \textsf{THEOREM2},   \textsf{AXIOM B}  using the definitions of $\mathsf{G}$ and $\mathsf{NE}$.
\item[\textsf{THEOREM 4}]  There exists a God-like being. 
   $$\boldsymbol{\exists}x\ \mathsf{G}(x)$$
   Follows from \textsf{THEOREM 3} together with \textsf{CORO} and  \textsf{AXIOM B}.
\end{description}

All claims have been verified with the higher-order proof assistant system Isabelle/HOL \parencite{Isabelle} and the sources of these
verification experiments are presented in Fig.~\ref{fig:ScottVariant} in the Appendix. This verification work utilised the universal meta-logical reasoning approach \parencite{J41} in order to obtain a ready to use ``implementation'' of higher-order modal logic in Isabelle/HOL's classical higher-order logic.


In these experiments only possibilist  quantifiers were  initially applied and later  the results were confirmed for a modified logical setting in which first-order actualist quantifiers for individuals were used, and otherwise possibilist quantifiers. It is also relevant to note that, in agreement with Gödel and Scott, in this article only intensions of (positive) properties paper are considered, in contrast to \textcite{fitting02:_types_tableaus_god}, who studied the use of extensions of properties in the context of the ontological argument. 


\section{Simplified Variant}\label{sec:SimplArgument}

Scott's (\citeyear{ScottNotes}) theory from above has interesting further corollaries, besides modal collapse $\mathsf{MC}$ and
monotheism (cf.~\cite{C40,C55}),\footnote{Monotheism results are of course dependent on the assumed notion of identity. This aspect should be further explored in future work.} and such corollaries can be explored using automated theorem proving technology.
In particular, the following two statements are implied.
\begin{description}
\item[\textsf{CORO 1}]  Self-difference 
is not a positive property. 
$$\boldsymbol {\neg} \mathsf{P}  (\lambda x\ (x \boldsymbol {\not=} x))$$
Since the setting in this article is extensional, we alternatively get that the empty property, $\lambda x\ \boldsymbol \bot$,  is not a positive property.
$$\boldsymbol {\neg} \mathsf{P}  (\lambda x\ \boldsymbol{\bot})$$
Both statements follow from \textsf{AXIOM 1} and \textsf{AXIOM 2}. This is easy to see, because if $\lambda x\ (x \boldsymbol {\not=} x)$ (respectively, $\lambda x\ \boldsymbol{\bot}$) was positive, then, by \textsf{AXIOM 2}, also its complement $\lambda x\ (x \boldsymbol {=} x)$ (respectively, $\lambda x\ \boldsymbol{\top}$) to be so, which contradicts \textsf{AXIOM 1}. Thus, only $\lambda x\ (x \boldsymbol{=} x)$ and $\lambda x\ \boldsymbol{\top}$ can be and indeed are positive, but not their complements.
\item[\textsf{CORO 2}]   A property is positive if it is entailed by a positive property.
 $$\boldsymbol{\forall} \phi\  \boldsymbol{\forall}  \psi\ ((\mathsf{P}(\phi) \boldsymbol{\wedge} (\boldsymbol{\forall} x\ (\phi(x)
                    \boldsymbol{\rightarrow} \psi(x)))) \boldsymbol{\rightarrow} \mathsf{P}(\psi))$$
  This follows from  \textsf{AXIOM 1} and \textsf{THEOREM 4} using the definition of  $\mathsf{G}$. Alternatively, the statement can be proved using  \textsf{AXIOM 1}, \textsf{AXIOM B} and modal collapse \textsf{MC}.       
\end{description}

The above observations are core motivation for our simplified variant of Gödel's argument as presented next; see \textcite{C85} for further experiments and explanations on the exploration on this and further simplified variants.\\

\begin{tcolorbox}[breakable,if odd page*={colback=gray!10}{colback=gray!10},
title={Axioms of the Simplified Ontological Argument}] 
\begin{description}
\item[\textsf{CORO 1}] Self-difference is not a positive property.
$$\boldsymbol {\neg} \mathsf{P}  (\lambda x\ (x \boldsymbol {\not=} x)) $$
{(Alternative: The empty property
  $\lambda x\ \boldsymbol \bot$ is not a positive property.)}
\item[\textsf{CORO 2}] A property entailed by a positive property is positive. 
$$\boldsymbol{\forall} \phi\  \boldsymbol{\forall}  \psi\ ((\mathsf{P}(\phi) \boldsymbol{\wedge} (\boldsymbol{\forall} x\ (\phi(x)
                    \boldsymbol{\rightarrow} \psi(x)))) \boldsymbol{\rightarrow} \mathsf{P}(\psi))$$
\item[\textsf{AXIOM 3}] Being Godlike is a positive property.
$$\mathsf{P}(\mathsf{G})$$
As before, an entity $x$ is defined to be God-like if it possesses all positive properties:
$$\mathsf{G}(x)  \quad \equiv  \quad \boldsymbol{\forall} \phi\ (\mathsf{P}(\phi)
\boldsymbol{\rightarrow} \phi(x))$$
\end{description}
\end{tcolorbox}

\vspace*{1em}

From the above axioms of the simplified theory the following successive argumentation steps can be derived in base modal logic K:
\begin{description}
\item[\textsf{LEMMA 1}] The existence of a non-exemplified positive property implies that self-difference (or, alternatively, the empty property) is a positive property.
    $$(\boldsymbol{\exists} \phi\  (\mathsf{P}(\phi) \boldsymbol{\wedge} \boldsymbol{\neg}\boldsymbol{\exists} x\ \phi(x))) 
        \boldsymbol{\rightarrow} \mathsf{P}  (\lambda x\ (x \boldsymbol {\not=} x))$$
        This follows from \textsf{CORO 2}, since such a $\phi$ would entail $\lambda x\ (x \boldsymbol {\not=} x)$.
\item[\textsf{LEMMA 2}] A non-exemplified positive property does not exist.
    $$\boldsymbol{\neg} \boldsymbol{\exists} \phi\  (\mathsf{P}(\phi) \boldsymbol{\wedge} \boldsymbol{\neg}\boldsymbol{\exists} x\ \phi(x)) $$
    Follows from \textsf{CORO 1} and the contrapositive of \textsf{LEMMA 1}.
\item[\textsf{LEMMA 3}] Positive properties are exemplified.
    $$\boldsymbol{\forall} \phi\  (\mathsf{P}(\phi) \boldsymbol{\rightarrow} \boldsymbol{\exists} x\ \phi(x)) $$
    This is just a reformulation of \textsf{LEMMA 2}.
\item[\textsf{THEOREM 3'}] There exists a God-like being.
  $$\boldsymbol{\exists}x\ \mathsf{G}(x)$$ 
  Follows from \textsf{AXIOM 3} and  \textsf{LEMMA 3}.
\item[\textsf{THEOREM 3}] Necessarily, there exists a God-like being.
  $$\boldsymbol{\Box} \boldsymbol{\exists}x\ \mathsf{G}(x)$$
  From \textsf{THEOREM 3'} by necessitation.
  \end{description}
  
The model finder \textit{nitpick} \cite{Nitpick} available in Isabelle/HOL can be employed to verify the consistency of this simple divine theory. The smallest satisfying model returned by the model finder consists of one possible world with one God-like entity, and with self-difference, resp.~the empty property, not being a positive property.
However, the model finder also tell us that it is impossible to prove \textsf{CORO}: $\boldsymbol{\Diamond} \boldsymbol{\exists}x\ \mathsf{G}(x)$, expressing that  the existence of a God-like being is possible. The simplest countermodel consists of a single possible world from which no other world is reachable, so that \textsf{CORO}, i.e.~$\boldsymbol{\Diamond} \boldsymbol{\exists}x\ \mathsf{G}(x)$, obviously cannot hold for this world, regardless of the truth of \textsf{THEOREM 3'}: $\boldsymbol{\exists}x\ \mathsf{G}(x)$ in it. However, the simple transition from the basic modal logic K to the logic KT eliminates this defect. To reach logic KT, \textsf{AXIOM T}: $ \boldsymbol{\forall} s\ (\boldsymbol{\Box} s \boldsymbol {\rightarrow} s)$  is postulated, that is, a property holds if it necessarily holds. This postulate appears uncontroversial. \textsf{AXIOM T} is equivalent to \textsf{AXIOM T'}: $ \boldsymbol{\forall} s\ (s \boldsymbol{\rightarrow} \boldsymbol{\Diamond} s)$, which expresses that a property that holds also possibly holds. Within modal logic KT we can thus obviously prove \textsf{CORO} from \textsf{THEOREM 3'} with the help of \textsf{AXIOM T'}.
                   
As an alternative to the above derivation of \textsf{THEOREM 3}, we can also proceed in logic KT analogously to the argument given in the introduction. 

\begin{description}
\item[\textsf{THEOREM 1}]  Positive properties are possibly exemplified. $$\boldsymbol{\forall} \phi\ 
                  (\mathsf{P}(\phi) \boldsymbol{\rightarrow}
                  \boldsymbol{\Diamond} \boldsymbol{\exists}x\ \phi(x))$$
                  Follows from \textsf{CORO 1}, \textsf{CORO 2} and \textsf{AXIOM T'}.
\item[\textsf{CORO}]  Possibly there exists a God-like being. $$\boldsymbol{\Diamond} \boldsymbol{\exists}x\ \mathsf{G}(x)$$
                  Follows from \textsf{THEOREM 1} and \textsf{AXIOM 3}.
\item[\textsf{THEOREM 2}]  The possible existence of a God-like being implies its necessary existence.
   $$ 
   \quad 
   \boldsymbol{\Diamond} \boldsymbol{\exists}x\ \mathsf{G}(x) \boldsymbol{\rightarrow} \boldsymbol{\Box} \boldsymbol{\exists}x\ \mathsf{G}(x) $$
     Follows from \textsf{AXIOM 3},  \textsf{CORO 1} and \textsf{CORO 2}.
\item[\textsf{THEOREM 3}]  Necessarily, there exists a God-like being. 
   $$\boldsymbol{\Box} \boldsymbol{\exists}x\ \mathsf{G}(x)$$
   Follows from \textsf{CORO} and \textsf{THEOREM2}.
\item[\textsf{THEOREM 3'}]  There exists a God-like being.
$$\boldsymbol{\exists}x\ \mathsf{G}(x)$$
   Follows from \textsf{THEOREM 3} with  \textsf{AXIOM T}.
\end{description}


Interestingly, the above simplified divine theory avoids modal collapse. This is confirmed by the model finder \textit{nitpick}, which reports a countermodel consisting of two possible worlds 
with one God-like entity.\footnote{In this countermodel, the possible worlds $i1$ and  $i2$ are reachable from $i2$, but only world $i1$ can be reached from $i1$. Moreover, there is non-positive property $\phi$ which holds for $e$ in world  $i2$ but not in $i1$. Apparently, in world $i2$, modal collapse $\boldsymbol{\forall} s\ (s \boldsymbol{\rightarrow} \boldsymbol{\Box} s)$ is not validated. The positive properties include $\lambda x\ \boldsymbol \top$.}

The above statements were all formally verified with Isabelle/HOL. As with Scott's variant,  only possibilist quantifiers were used initially, and later the results were confirmed also for a modified logical setting in which first-order actualist quantifiers for individuals were used, and possibilist quantifiers otherwise. The Isabelle/HOL sources of the conducted verification studies are presented in Figs.~\ref{fig:HOML}-\ref{fig:SimpleVariantActualist} in the Appendix. 

In the related exploratory studies \parencite{C85}, a suitably adapted notion of a modal ultrafilter was additionally used to support the comparative analysis of different variants of Gödel's ontological argument, including those proposed by \textcite{AndersonGettings} and \textcite{fitting02:_types_tableaus_god}, which avoid modal collapse. 
These experiments are a good demonstration of the maturity that modern theorem proving systems have reached. These systems are ready to fruitfully support the exploration of metaphysical theories.

The development of Gödel's ontological argument has recently been addressed by 
\textcite{KanckosLethen19}. They discovered previously unknown variants
of the argument in Gödel's Nachlass, whose relation to the presented simplified variants
should be further investigated in future work.
The version No.~2 they reported has meanwhile been
 formalised and verified in Isabelle/HOL, similar to the work presented above.
This version No.~2  avoids the
  notions of essence and necessary existence and associated
  definitions/axioms, just as our simplified version does. However, this version, in many respects, also differs from ours, and it assumes a higher-modal modal logic S5. 
  
 Further variants whose relation to the presented simplified  argument should be studied in further work include \textcite{Gustafsson19} and \textcite{Swietorzecka20}, respectively \textcite{Christian89}.
 
 In future work, I would also like to  further deepen ongoing studies of Fitting's (\citeyear{fitting02:_types_tableaus_god}) proposal, which focuses with extensions rather than intensions of (positive) properties.


\section{Discussion}\label{sec:discussion}
Whether the simplified variant of Gödel's ontological argument presented in this paper actually increases or decreases the argument's appeal and persuasiveness is a question I would like to pass on to philosophy and theology. As a logician, I see my role primarily as providing useful input and clarity to promote informed debate. 


I have shown how a significantly simplified version of Gödel's ontological variant can be explored and verified in interaction with modern theorem proving technology. Most importantly, this simplified variant avoids modal collapse, and some further issues, which have triggered criticism on Gödel's argument in the past. Future work could investigate the extent to which such theory simplification studies could even be fully automated. The resulting rational reconstructions of argument variants would be very useful in gaining more intuition and understanding of the theory in question, in this case a theistic theory, which in turn could lead to its demystification and also to the identification of flawed discussions in the existing literature.

\paragraph{Acknowledgements:}  I thank Andrea Vestrucci  and the anonymous reviewers for valuable comments that helped improve this article.

\printbibliography

\vskip5em

\clearpage
\pagebreak
\section*{Appendix: Sources of Conducted Experiments}
\addcontentsline{toc}{section}{Appendix: Sources of Conducted Experiments}
\begin{figure}[h] \centering
\fcolorbox{gray!10}{gray!30}{\includegraphics[width=.9\columnwidth]{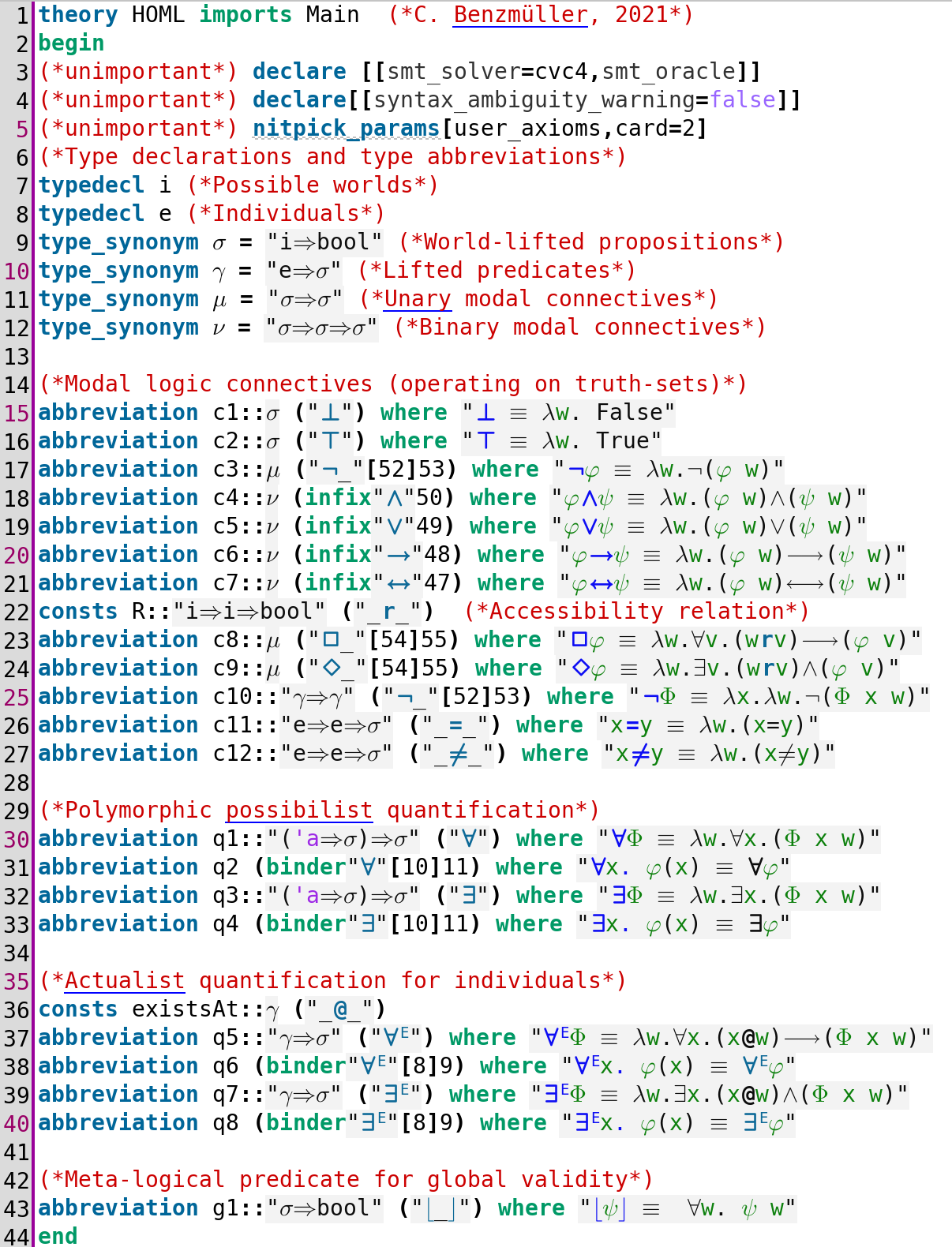}}
\caption{The universal meta-logical reasoning approach at work: exemplary shallow semantic embedding of modal higher-order logic K in classical higher-order logic. \label{fig:HOML}}
\end{figure}

\begin{figure}[t] \centering
\fcolorbox{gray!10}{gray!30}{\includegraphics[width=.9\columnwidth]{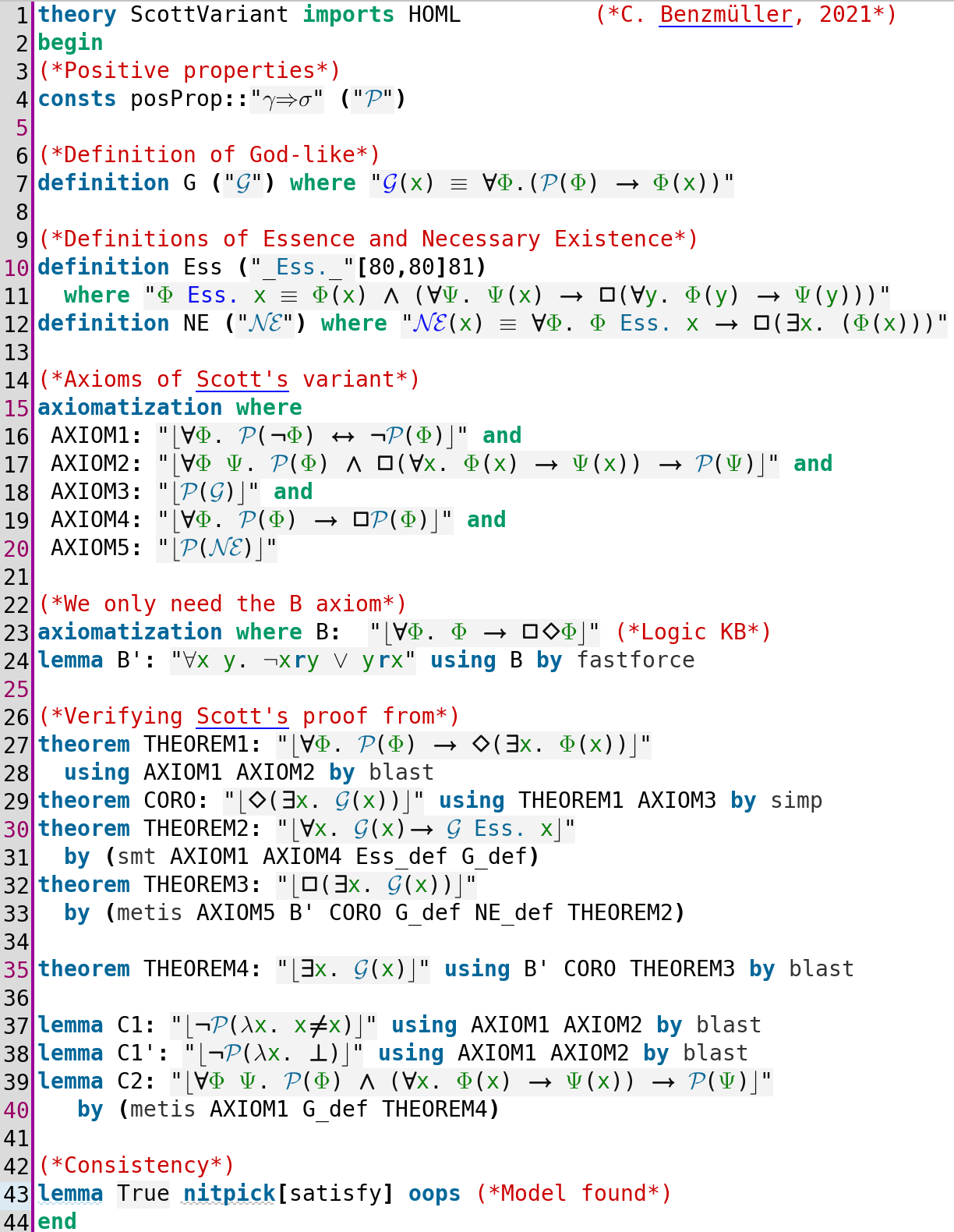}}
\caption{Verification of Scott's variant of Gödel's ontological argument in modal higher-order logic KB, using first-order and higher -order possibilistic quantifiers; the theory HOML from Fig.~\ref{fig:HOML} is imported.  \label{fig:ScottVariant}}
\end{figure}

\begin{figure}[t] \centering
\fcolorbox{gray!10}{gray!30}{\includegraphics[width=.9\columnwidth]{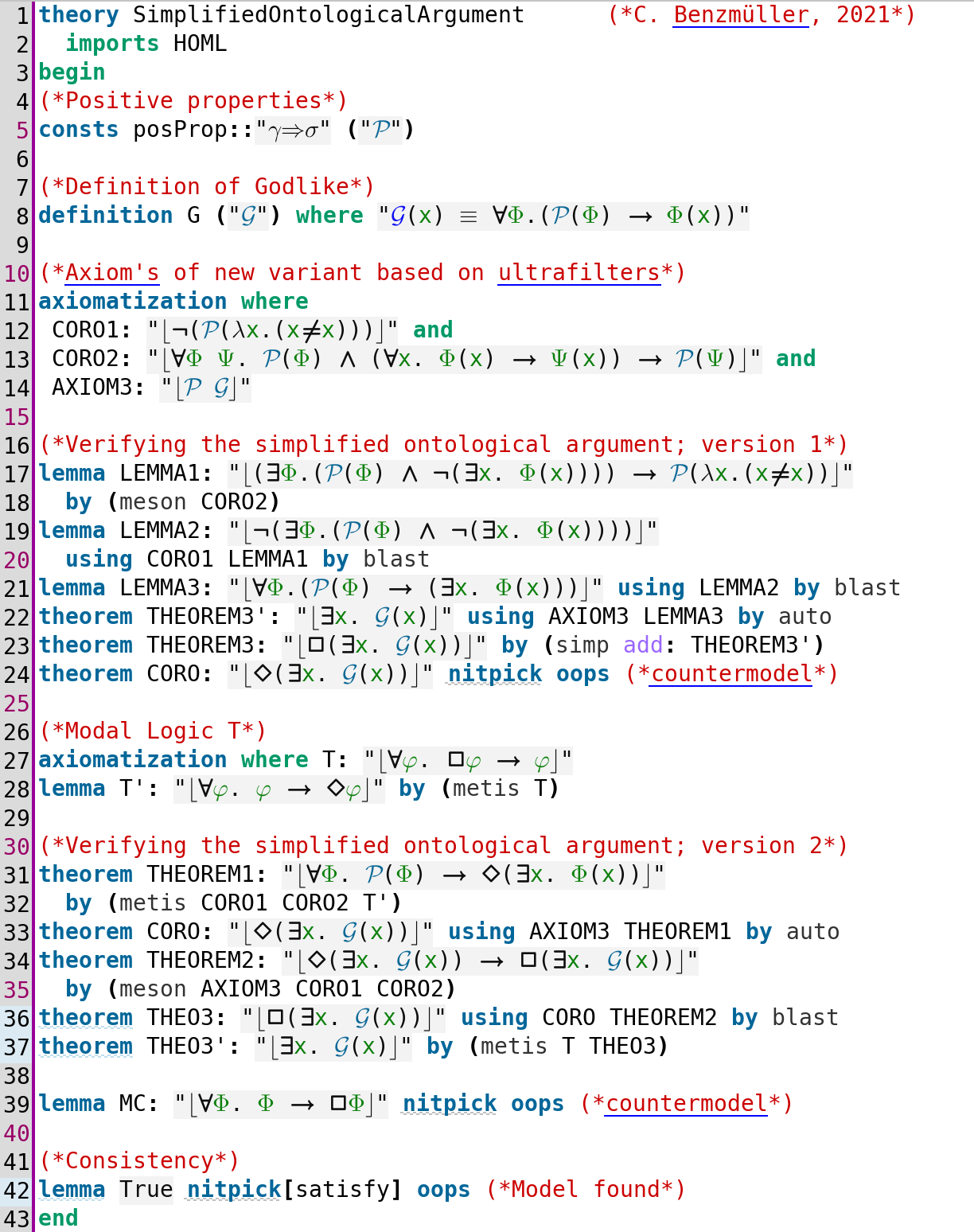}}
\caption{Simplified ontological argument in modal logic K, respectively KT, using possibilist first-order and higher-order quantifiers. \label{fig:SimpleVariant}}
\end{figure}

\begin{figure}[t] \centering
\fcolorbox{gray!10}{gray!30}{\includegraphics[width=.9\columnwidth]{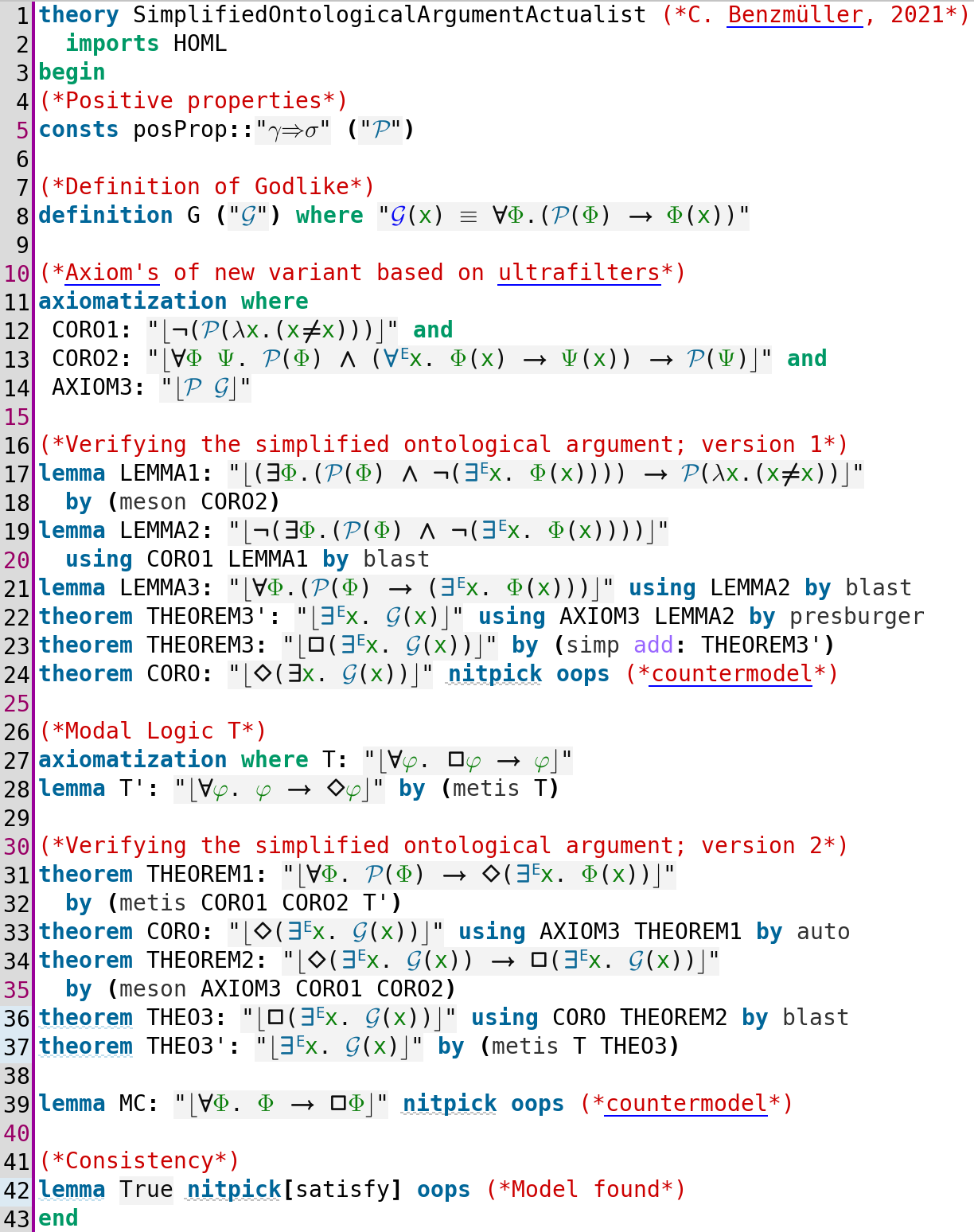}}
\caption{Simplified ontological argument in modal logic K, respectively KT, using actualist quantifiers first-order quantifiers and possibilist higher-order quantifiers. \label{fig:SimpleVariantActualist}}
\end{figure}

\end{document}